\documentclass[aps,prd,reprint,nofootinbib,superscriptaddress]{revtex4-2}

\usepackage{amsfonts}
\usepackage{amsmath}
\usepackage{amssymb}
\usepackage{mathtools}
\usepackage{xcolor}

\usepackage{bm}
\usepackage{graphicx}
\usepackage{fancybox}
\usepackage{comment}
\usepackage{multirow}
\usepackage{tipa}
\definecolor{refs}{RGB}{245,156,74}
\usepackage[colorlinks=true,hyperfootnotes=true,citecolor=cyan]{hyperref}
\usepackage{enumitem}
\usepackage{comment}
\usepackage{subcaption}
\usepackage{esint}
\usepackage{bbold}
\usepackage{csquotes}
\usepackage{slashed}

\usepackage{tikz}
\usetikzlibrary{arrows.meta, positioning, shapes.geometric, decorations.pathreplacing}

\newcommand{\be}{\begin{equation}}
\newcommand{\ee}{\end{equation}}
\newcommand{\ba}{\begin{eqnarray}}
\newcommand{\ea}{\end{eqnarray}}
\newcommand{\bs}{\begin{subequations}}
\newcommand{\es}{\end{subequations}}

\newcommand{\e}{\mathrm{e}}
\newcommand{\ie}{\text{\textschwa}}

\newcommand{\diff}{\textrm{d}}

\newcommand{\xx}{\hat{x}}

\newcommand{\lp}{\left(}
\newcommand{\rp}{\right)}
\newcommand{\lb}{\left[}
\newcommand{\rb}{\right]}

\newcommand{\nn}{\nonumber}

\newcommand{\x}{ \prescript{\pm}{}}

\DeclareMathSymbol{\mrq}{\mathord}{operators}{`'}

\usepackage{graphicx}

\begin{document}

\title{Canonical currents without improvements: the QCD example}

\author{Tomi Koivisto}
\email{tomi.koivisto@ut.ee}
\affiliation{Laboratory of Theoretical Physics, Institute of Physics, University of Tartu, W. Ostwaldi 1, 50411 Tartu, Estonia}
\affiliation{National Institute of Chemical Physics and Biophysics, R\"avala pst. 10, 10143 Tallinn, Estonia}

\begin{abstract}
In conventional formulations of relativistic field theory, energy-momentum and angular-momentum currents are defined only up to
improvement terms in flat spacetime, and in curved spacetime they are conserved only along symmetries of the metric. When defined
as canonical Noether currents relative to an independent parallel frame, as in General Parallel Relativity, they are both unambiguous
and conserved. For QCD, the gauge-invariant tensor, the Belinfante--Rosenfeld relation and the spin surface term follow from the
variational principle. An exact coupled solution is shown to have conserved energy and rotation charges without the corresponding
symmetries. Quarks of opposite helicity are indistinguishable by the metric and the symmetric energy-momentum tensor, yet carry
opposite rotation charges.
\end{abstract}

\maketitle

\section{Introduction}

The distinction between canonical and metric energy-momentum is commonly addressed by adding improvement terms to the former
\cite{1940Phy.....7..449B,Rosenfeld1940}. These terms have recently been revisited through gauge invariance
\cite{Brauner:2019lcb,Kourkoulou:2022ajr,Haberzettl:2024,Hobson:2026vkg} and Noether's second theorem \cite{Freese:2025glz,Singh:2024}.
Metric-affine geometry relates them to the response of matter to independent connections
\cite{Hecht:1992xn,Iosifidis:2025sjx,Iosifidis:2026heu}. Although this illuminates familiar tensor identities, it does not by itself
identify the physical energy-momentum current, an ambiguity that also underlies debates on spin decompositions
\cite{Ji:1996ek,Lorce:2017wkb} and on pseudo-gauges in spin hydrodynamics \cite{Speranza:2020ilk,Huang:2024}. With gravity, the problem is worse,
since conservation then requires spacetime symmetries \cite{Szabados:2009eka}, and even linearized gravity admits no preferred local
gauge-invariant energy-momentum tensor \cite{Hobson:2026vkg,Hobson:2026lin}.

General Parallel Relativity (G$_\parallel$R) addresses both problems \cite{Koivisto:2018aip,BeltranJimenez:2019odq,Gomes:2023hyk,Koivisto:2026grg}.
Energy and momentum are canonical Noether quantities defined relative to a parallel frame, and their frame dependence expresses
relativity rather than ambiguity. In a canonical frame, which itself carries no energy-momentum, the matter currents are conserved on
shell. The existence of such frames provides a covariant interpretation of the equivalence principle.

We illustrate this in QCD. We show that (i) colour-compensated Noether currents yield the gauge-invariant tensor and its
Belinfante--Rosenfeld relation directly from the action, (ii) in the canonical frame these currents are conserved without Killing
vectors, and (iii) rotation charges in finite regions carry spin information that the symmetric tensor does not. An exact coupled
solution shows (i)--(iii) at work.

\section{The coupled variational principle}

The basic gravitational fields of G$_\parallel$R are the metric tetrad and the reference coframe.\footnote{Indices $a,b,\dots$ label
the $GL$ frame, $\mu,\nu,\dots$ spacetime coordinates, and $I,J,\dots$ metric Lorentz indices. $\sum_f$ runs over quark flavours.
$\kappa=(8\pi G)^{-1}$, $g_s$ is the strong coupling, and $\eta_{IJ}=\text{diag}(-1,1,1,1)$. Weight-$1/2$ (anti)symmetrization is
used throughout.}
Let $\e^a{}_\mu$ be a reference coframe with dual $\ie_a{}^\mu$, independent of the metric tetrad $h^I{}_\mu$ with dual $h_I{}^\mu$.
The latter determines the metric $g_{\mu\nu}=\eta_{IJ}h^I{}_\mu h^J{}_\nu$, its Christoffel symbols $\mathring{\Gamma}^\alpha{}_{\mu\nu}$
and the Levi-Civita derivative $\mathring{\nabla}_\mu$. The reference frame defines the symmetric connection
$\Gamma^\alpha{}_{\mu\nu}=\ie_a{}^\alpha\partial_{(\mu}\e^a{}_{\nu)}$ and thereby an independent torsion-free derivative $\nabla_\mu$.
We use $\nabla$ to couple matter to gravity, but since $\Gamma$ is torsion free, the conventional choice $\mathring{\nabla}$ could
equivalently be made \cite{BeltranJimenez:2020sih}.

The matter fields are the quarks $(\psi_f,\bar{\psi}_f)$ and the colour connection $A_\mu=A^A_\mu t_A$, with
$\text{tr}(t_At_B)=\delta_{AB}/2$. Since $A$ is a connection, its curvature is
\be \label{F}
F_{\mu\nu}=2\partial_{[\mu}A_{\nu]}-ig_s[A_\mu,A_\nu]\,,
\ee
independent of $\Gamma$. Here and below, $\nabla$ and $\mathring{\nabla}$ include the colour connection, acting on $A_\nu$ and
$F_{\mu\nu}$ in the adjoint. The spinor covariant derivative is
\be
\nabla_\mu\psi=\lb\partial_\mu-\tfrac{1}{4}h_{I\nu}\lp\partial_\mu h_J{}^\nu+\Gamma^\nu{}_{\mu\rho}h_J{}^\rho\rp\gamma^{IJ}-ig_sA_\mu\rb\psi\,,
\ee
with $\gamma^{(I}\gamma^{J)}=-\eta^{IJ}$, $\gamma^{IJ}=\gamma^{[I}\gamma^{J]}$, $\gamma^\mu=\gamma^Ih_I{}^\mu$ and
$\bar{\psi}=\psi^\dagger\gamma^0$. The derivative $\mathring{\nabla}_\mu\psi$ is defined by replacing $\Gamma\to\mathring{\Gamma}$. The action of the
coupled theory is
\ba \label{action}
I & = & \int\diff^4x\sqrt{-g}\Big[-\kappa X^\alpha{}_{[\beta}{}^\beta X^\gamma{}_{\gamma]\alpha}-\tfrac{1}{2}\text{tr}\lp F_{\mu\nu}F^{\mu\nu}\rp \nn \\
& + & \sum_f\lp\tfrac{i}{2}\bar{\psi}_f\gamma^\mu\overset{\leftrightarrow}{\nabla}_\mu\psi_f-m_f\bar{\psi}_f\psi_f\rp\Big]\,,
\ea
where
\be \label{X}
X^\alpha{}_{\mu\nu}=\ie_a{}^\alpha\partial_\mu\e^a{}_\nu-\mathring{\Gamma}^\alpha{}_{\mu\nu}\,,
\ee
and $\bar{\psi}\gamma^\mu\overset{\leftrightarrow}{\nabla}_\mu\psi\equiv\bar{\psi}\gamma^\mu\nabla_\mu\psi-(\nabla_\mu\bar{\psi})\gamma^\mu\psi$.
The gravitational term equals $\kappa\mathring{R}/2$ up to a divergence. The QCD action $I_{\text{QCD}}$, the part of (\ref{action})
isolated by setting $\kappa=0$, depends on $h$ but not on $\e$.

Its symmetric Hilbert energy-momentum tensor is
\be \label{hilbert}
T^{\mu\nu}=\tau_I{}^{(\mu}h_J{}^{\nu)}\eta^{IJ}\,, \quad
\tau_I{}^\mu\equiv\frac{1}{\sqrt{-g}}\frac{\delta I_{\text{QCD}}}{\delta h^I{}_\mu}\,.
\ee
With spinors on shell, local Lorentz invariance makes $\tau_I{}^\mu h^{I\nu}$ symmetric and the explicit symmetrization is not
required. Variation of the matter action and integration by parts yield the boundary term $\partial_\mu\theta^\mu_{\text{QCD}}$, with
\ba
\theta^\mu_{\text{QCD}}[\delta] & = & -\sqrt{-g}\,F^{\mu\nu}_A\delta A^A_\nu+\tfrac{1}{2}\mathcal{S}^{\mu IJ}h_{J\nu}\delta h_I{}^\nu \nn \\
& + & \tfrac{i}{2}\sqrt{-g}\sum_f\lp\bar{\psi}_f\gamma^\mu\delta\psi_f-\delta\bar{\psi}_f\gamma^\mu\psi_f\rp\,,
\ea
where the quark spin tensor density
\be
\mathcal{S}^{\rho\mu\nu}=\tfrac{i}{4}\sqrt{-g}\sum_f\bar{\psi}_f\{\gamma^\rho,\gamma^{\mu\nu}\}\psi_f
\ee
is completely antisymmetric. The spin contribution thus comes directly from the action, without an improvement
prescription or an embedding in a metric-affine geometry.

\section{The matter Noether currents}

Consider the Bessel-Hagen transformations \cite{BesselHagen1921,Haberzettl:2024,Hobson:2026vkg}, a diffeomorphism generated by
$\xi^\mu$ combined with colour and Lorentz compensation,
\be \label{BH}
\hat{\delta}_\xi A_\mu=\xi^\nu F_{\nu\mu}\,, \quad \hat{\delta}_\xi\psi=\xi^\mu\mathring{\nabla}_\mu\psi\,, \quad
\hat{\delta}_\xi h^I{}_\mu=h^I{}_\nu\mathring{\nabla}_\mu\xi^\nu\,.
\ee
The metric derivative pairs the spinor and tetrad transformations consistently, with
$\hat{\delta}_\xi g_{\mu\nu}=2\mathring{\nabla}_{(\mu}\xi_{\nu)}$. The Dirac equations give
$\frac{i}{2}\bar{\psi}_f\gamma^\mu\overset{\leftrightarrow}{\mathring{\nabla}}_\mu\psi_f=m_f\bar{\psi}_f\psi_f$, so each quark
Lagrangian vanishes on shell, without a massless approximation. The action then gives the Noether current density
\cite{Noether:1918zz,Koivisto:2026surface}
\ba \label{current}
J^\mu_\xi & = & \sqrt{-g}\,\xi^\mu\mathcal{L}_{\text{QCD}}-\theta^\mu_{\text{QCD}}[\hat{\delta}_\xi] \nn \\
& = & \Theta^{\mu\nu}\xi_\nu-\tfrac{1}{2}\mathcal{S}^{\lambda\mu\nu}\mathring{\nabla}_\lambda\xi_\nu\,,
\ea
where
\ba \label{Theta}
\Theta^{\mu\nu} & = & \sqrt{-g}\Big[2\,\text{tr}\lp F^{\mu\lambda}F^\nu{}_\lambda-\tfrac{1}{4}g^{\mu\nu}F_{\rho\sigma}F^{\rho\sigma}\rp \nn \\
& - & \tfrac{i}{2}\sum_f\bar{\psi}_f\gamma^\mu\overset{\leftrightarrow}{\mathring{\nabla}}{}^\nu\psi_f\Big]\,.
\ea
Colour compensation gives the gauge-invariant tensor density $\Theta^{\mu\nu}$, replacing the gauge-dependent tensor obtained from
uncompensated translations \cite{Freese:2025glz,Brauner:2019lcb,Kourkoulou:2022ajr}. It is independent of the Lorentz lift, since the
spinor and tetrad contributions cancel for pure Lorentz transformations \cite{Kim:2024ewt}.

The Noether identity of local Lorentz invariance reads, on the Dirac equations \cite{1940Phy.....7..449B,Hehl:1976kj},
\be \label{lorentz}
\partial_\lambda\mathcal{S}^{\lambda\mu\nu}=\Theta^{\nu\mu}-\Theta^{\mu\nu}\,,
\ee
where $\mathring{\nabla}_\lambda\mathcal{S}^{\lambda\mu\nu}=\partial_\lambda\mathcal{S}^{\lambda\mu\nu}$ for a totally antisymmetric
density.
Together with the symmetry of (\ref{hilbert}) it yields
\be \label{BR}
\sqrt{-g}\,T^{\mu\nu}=\Theta^{(\mu\nu)}=\Theta^{\mu\nu}+\tfrac{1}{2}\partial_\lambda\mathcal{S}^{\lambda\mu\nu}\,,
\ee
and hence
\be \label{currentT}
J^\mu_\xi=\sqrt{-g}\,T^{\mu\nu}\xi_\nu-\tfrac{1}{2}\partial_\lambda\lp\mathcal{S}^{\lambda\mu\nu}\xi_\nu\rp\,.
\ee
Thus the Belinfante--Rosenfeld relation for the gauge-compensated tensor, whose symmetric part defines the
gravitational form factors of hadrons \cite{Polyakov:2018zvc,Burkert:2018bqq}, is obtained directly from the matter action.\footnote{In
the free flat Dirac limit the spin divergence agrees with Ref.~\cite{Iosifidis:2025sjx}, and the gluon part of (\ref{Theta}) is
already symmetric by colour compensation. In metric-affine constructions \cite{Hecht:1992xn,Iosifidis:2025sjx,Iosifidis:2026heu} the
improvement arises from the response to an independent connection, which depends on how the matter action is extended. For gauge
fields the curvature (\ref{F}) is independent of $\Gamma$, so a nonzero response requires either a gauge-breaking extension or a
nonminimal coupling that changes the theory.}

For an arbitrary generator $\xi^\mu$ the matter equations give the balance law
\be \label{balance}
\partial_\mu J^\mu_\xi=\frac{\delta I_{\text{QCD}}}{\delta h^I{}_\rho}\hat{\delta}_\xi h^I{}_\rho
=\sqrt{-g}\,T^{\mu\nu}\mathring{\nabla}_{(\mu}\xi_{\nu)}\,.
\ee
In the special case that $\xi^\mu$ is a Killing vector of the metric, the corresponding matter current is conserved. Conventional
formulations stop here.

\section{The canonical parallel frame}

The conservation of the canonical energy and momentum currents, however, is independent of the metric configuration. Consider
translations along the reference frame, $\xi^\mu=\ie_a{}^\mu$, and write $J^\mu_a\equiv J^\mu_{\ie_a}$. The gravitational part of
(\ref{action}) has its own Noether current $J^\mu_{G,a}$, the canonical energy-momentum of the frame, related to the metric curvature
by the Noether identity \cite{Koivisto:2026grg}
\be \label{JG}
J^\mu_{G,a}=-\kappa\sqrt{-g}\,\mathring{G}^\mu{}_\nu\ie_a{}^\nu+\partial_\lambda H_a{}^{\lambda\mu}\,,
\ee
with the antisymmetric superpotential
\be \label{superpotential}
H_a{}^{\lambda\mu}=-\kappa\sqrt{-g}\,\ie_a{}^\alpha\lp X^{[\lambda}{}_\alpha{}^{\mu]}+2\delta^{[\lambda}_\alpha X^{\mu]}\rp\,,
\ee
where $X^\mu\equiv X^{[\mu}{}_\alpha{}^{\alpha]}$. The total current $J^\mu_{\text{tot},\xi}=J^\mu_\xi+J^\mu_{G,\xi}$ obeys
\be
\partial_\mu J^\mu_{\text{tot},\xi}=\sqrt{-g}\lp T^{\mu\nu}-\kappa\mathring{G}^{\mu\nu}\rp\mathring{\nabla}_{(\mu}\xi_{\nu)}\,,
\ee
so the coupled field equation $\kappa\mathring{G}^{\mu\nu}=T^{\mu\nu}$ conserves the total current for every $\xi$. It does not make
the matter current separately conserved.

This is where the equivalence principle enters. By relativity, the physical reference translations are those satisfying the
canonical condition
\be \label{canonical}
J^\mu_{G,a}=0\,,
\ee
which states that the reference frame itself carries no energy-momentum. In a \emph{canonical frame} \cite{BeltranJimenez:2019odq} the
matter translation currents coincide with the conserved total currents. Equations (\ref{currentT}) and (\ref{JG}), with the field
equations, give their explicit surface representation,
\ba \label{surface}
& & J^\mu_a=J^\mu_{\text{tot},a}\,, \quad \partial_\mu J^\mu_a=0\,, \nn \\
& & J^\mu_a=\partial_\lambda\lp H_a{}^{\lambda\mu}-\tfrac{1}{2}\mathcal{S}^{\lambda\mu\nu}\ie_{a\nu}\rp\,.
\ea
Energy and momentum are therefore conserved on any coupled solution admitting a canonical reference, whether or not the metric has
isometries. No holonomic-frame assumption is needed.

Two objections can be anticipated. First, that a Noether current is on shell the curl of a superpotential follows from Noether's
second theorem \cite{Noether:1918zz}, recently invoked for the QCD tensor and for free Dirac currents
\cite{Freese:2025glz,Singh:2024}. Here the superpotential follows from the coupled action with gravity and gauge fields, and the
canonical frame selects which surface charges are physical. Second, Eq.~(\ref{canonical}) does not follow from varying the action,
since $\e$ has no bulk field equation. It is the defining equation of the canonical frames, generalizing the special-relativistic
inertial frames, which are likewise physical but not determined by the variational field equations.

\section{Rotation charges}

A rotation is not a constant linear combination of parallel translations. Write an arbitrary generator as $\xi=f^a\ie_a$, with
$f^a=\e^a{}_\nu\xi^\nu$. In the independent variables $(\e,h)$, the derivatives $\partial_\lambda f^a$ enter the boundary potential
through both the reference coframe and the metric tetrad, and their combined coefficient is the superpotential
(\ref{superpotential}). Hence
\be
J^\mu_{G,\xi}=f^aJ^\mu_{G,a}+(\partial_\lambda f^a)H_a{}^{\lambda\mu}\,.
\ee
The canonical condition removes the first term, not the second. In any reference, Eqs.~(\ref{currentT}) and (\ref{JG}) give the total
current on the matter equations,
\ba \label{Q}
J^\mu_{\text{tot},\xi} & = & \sqrt{-g}\lp T^{\mu\nu}-\kappa\mathring{G}^{\mu\nu}\rp\xi_\nu+\partial_\lambda Q_\xi^{\lambda\mu}\,, \nn \\
Q_\xi^{\lambda\mu} & = & f^aH_a{}^{\lambda\mu}-\tfrac{1}{2}\mathcal{S}^{\lambda\mu\nu}\xi_\nu\,.
\ea
On coupled solutions $J^\mu_{\text{tot},\xi}=\partial_\lambda Q^{\lambda\mu}_\xi$. The matter rotation charge is therefore not
conserved merely because the translation charges are. What G$_\parallel$R provides is the surface potential $Q_\xi$ and the exchange of
angular momentum between matter and gravity in the physical frame \cite{Koivisto:2026surface}.

For a rotation $\varphi$ of period $2\pi$, tangent to a fixed closed two-surface $\mathcal{B}$, let $Q_\varphi$ be the two-form
associated with (\ref{Q}). On a phase space whose boundary data the rotation preserves, its Hamiltonian variation is
\cite{Koivisto:2026surface,Iyer:1994ys}
\be \label{hamiltonian}
\delta H_\varphi=\oint_{\mathcal{B}}\lb\delta Q_\varphi+i_\varphi\lp\theta_G+\theta_{\text{QCD}}\rp\rb=\delta\oint_{\mathcal{B}}Q_\varphi\,.
\ee
The last equality is geometric, since a three-form evaluated on $\varphi$ and two vectors tangent to $\mathcal{B}$ vanishes. Tangency
thus ensures integrability without a Killing vector in the bulk. In a finite region, conservation further requires that no symplectic flux crosses the
boundary, as we verify for the example below.

In the flat limit, $g_{\mu\nu}\to\eta_{\mu\nu}$ and $\e^a\to\diff x^a$, Eq.~(\ref{current}) yields the kinetic angular-momentum current
$J^{\rho\mu\nu}_{\text{kin}}=2x^{[\mu}\Theta^{|\rho|\nu]}+\mathcal{S}^{\rho\mu\nu}$ of Ji's gauge-invariant decomposition
\cite{Ji:1996ek}, and for a finite region $V$ the kinetic and Belinfante angular momenta differ by a spin surface term
\cite{Lorce:2017wkb},
\be
J^{\mu\nu}_{\text{kin}}(V)=J^{\mu\nu}_{\text{BR}}(V)-\tfrac{1}{2}\oint_{\partial V}\diff S\,\hat{n}_k\lp x^\mu\mathcal{S}^{k0\nu}-x^\nu\mathcal{S}^{k0\mu}\rp\,.
\ee
By complete antisymmetry, $s_\lambda n_\mu\mathcal{S}^{\lambda\mu\nu}\xi_\nu$ on a boundary with unit normals $n,s$ picks out only the
part of $\xi$ tangent to the two-surface. Normal time evolution therefore receives no spin correction, while rotations can. In
G$_\parallel$R this statement survives full gravitational coupling, as we now show.

\begin{figure}[t]
\centering
\begin{tikzpicture}[scale=0.9, every node/.style={font=\small}]
  \node at (-0.4,2.3) {(a)};
  \draw[thick] (0.6,0) ellipse (0.35 and 1.0);
  \draw[thick] (0.6,1.0) -- (6.0,1.0);
  \draw[thick] (0.6,-1.0) -- (6.0,-1.0);
  \draw[thick] (6.0,1.0) arc (90:-90:0.35 and 1.0);
  \draw[thick, dashed, gray] (6.0,1.0) arc (90:270:0.35 and 1.0);
  \draw[thick, blue] (5.2,1.0) arc (90:-90:0.35 and 1.0);
  \node[blue] at (5.85,-0.35) {$\varphi$};
  \draw[teal, thick, domain=1.0:5.0, samples=120, smooth] plot (\x, {0.35*sin(deg(3.3*\x))});
  \node[teal] at (1.8,-0.6) {gluon wave};
  \foreach \xx in {1.6,2.6,3.6,4.6} \draw[red!80!black, -{Stealth}, thick] (\xx-0.3,0.15) -- (\xx+0.3,0.15);
  \node[red!80!black] at (3.6,0.6) {quark spin};
  \draw[-{Stealth}, thick] (6.6,0) -- (7.4,0) node[right] {$k$};
  \draw[-{Stealth}, thick] (3.3,1.0) -- (3.3,1.7) node[above] {$s$};
  \draw[{Stealth}-{Stealth}] (0.0,0) -- (0.0,1.0); \node at (-0.25,0.5) {$R$};
  \draw[-{Stealth}, gray] (0.8,-1.25) to[bend right=8] (5.8,-1.25);
  \node at (3.3,-1.65) {$z\sim z+L$};
  \node at (-0.4,-2.4) {(b)};
  \node at (2.6,-2.5) {gluon $+$}; \node at (4.9,-2.5) {gluon $-$};
  \node at (0.7,-3.1) {quark $+$}; \node at (0.7,-3.8) {quark $-$};
  \node[draw, fill=red!8, minimum width=2.0cm] at (2.6,-3.1) {$H_\varphi=+N_q/2$};
  \node[draw, fill=red!8, minimum width=2.0cm] at (4.9,-3.1) {$H_\varphi=+N_q/2$};
  \node[draw, fill=blue!8, minimum width=2.0cm] at (2.6,-3.8) {$H_\varphi=-N_q/2$};
  \node[draw, fill=blue!8, minimum width=2.0cm] at (4.9,-3.8) {$H_\varphi=-N_q/2$};
  \draw[decorate, decoration={brace, mirror, amplitude=5pt}] (1.5,-4.2) -- (6.0,-4.2);
  \node at (3.75,-4.7) {same $g_{\mu\nu}$, $T_{\mu\nu}$};
\end{tikzpicture}
\caption{(a) The finite cell $V=D^2_R\times S^1_L$. A quark pulse and a circularly polarized gluon wave propagate along $k$. The
rotation $\varphi$ is tangent to the boundary, whose outward normal is $s$. (b) The rotation charge (\ref{Hphi}) is determined solely by the quark helicity.}
\label{fig:cell}
\end{figure}
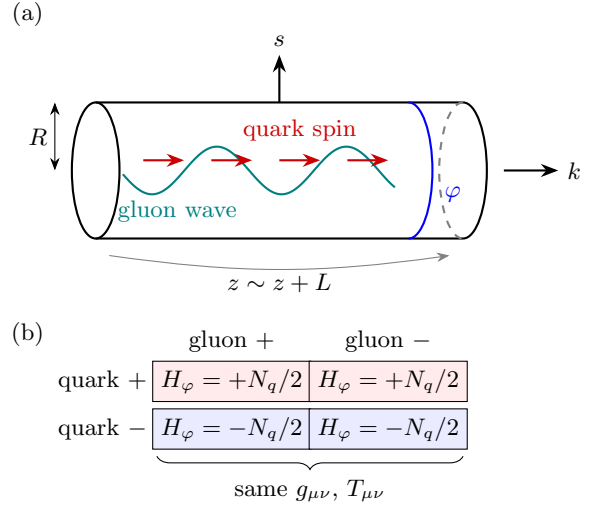

\section{Quarks and gluons on a gravitational wave}
\label{sec:example}

Let us construct an exact coupled solution illustrating (ii) and (iii): the canonical energy and rotation charges are
conserved without the corresponding isometries, and the rotation charge distinguishes configurations
that the metric cannot. A simple setting is radiation. For null fields aligned with a covariantly constant null
vector $k$, the energy-momentum tensor is proportional to $k_\mu k_\nu$, and the Einstein equations reduce to a Poisson equation in
the transverse plane. The quarks are massless with spin along $k$, the gluons form a plane wave along $k$, and a vacuum
gravitational wave may cross both. For the gluon wave to remain free, the quark colour current must vanish, which we arrange with one
flavour per colour.

With $u=t-z$ and $k_\mu\diff x^\mu=\diff u$, a pp-wave ansatz familiar from Einstein--Weyl theory \cite{Cianci:2015pba} solves all
field equations. The reference frame is Cartesian and the metric is a pp-wave,
\bs \label{solution}
\be \label{sol-geometry}
\e^a=\diff x^a\,, \quad g_{\mu\nu}=\eta_{\mu\nu}+Wk_\mu k_\nu\,.
\ee
Three massless flavours, whose colour vectors $c_f$ are orthonormal eigenvectors of $t_3$, $t_3c_f=\lambda_fc_f$ with
$\sum_f\lambda_f=0$, form a polarized pulse,
\be \label{sol-quarks}
\psi_f=\sqrt{\tfrac{n}{3}}\,e^{-i\Phi}\chi\otimes c_f\,, \quad \lp\gamma^{\hat{0}}-\gamma^{\hat{3}}\rp\chi=0\,, \quad \chi^\dagger\chi=1\,,
\ee
of total number density $n(u)>0$ and phase $\Phi(u)$, so that each quark carries null momentum along $k$
with energy $\Phi'>0$. The gluons form a circularly polarized wave of helicity $\pm$,
\be \label{sol-gluons}
A=-t_3\,x^i\varepsilon_i\,\diff u\,, \quad \varepsilon=E_0\lp\cos\omega_gu,\pm\sin\omega_gu\rp\,,
\ee
with transverse field strength $F=t_3\,k\wedge\varepsilon$, $F_{ui}=t_3\varepsilon_i$.\footnote{Equivalently, $A_u=0$ and
$A_i=t_3\int^u\varepsilon_i(u')\,\diff u'$ in the transverse gauge.} Finally the Einstein equations fix the wave profile,
\be \label{sol-W}
W=-\frac{\rho+E_0^2}{2\kappa}\lp x^2+y^2\rp+b_+\lp x^2-y^2\rp+2b_\times xy\,,
\ee
\es
where $\rho=\Phi'n$ is the quark energy density. The functions $n$, $\Phi$, $b_+$ and $b_\times$ of $u$ are arbitrary, and $b_+,b_\times$ describe the vacuum
gravitational wave. The solution contains two independent helicity choices, the quark helicity
$\Sigma_3\chi_\pm=\pm\chi_\pm$ with $\Sigma_3=i\gamma^{\hat{1}}\gamma^{\hat{2}}$, and the sign in (\ref{sol-gluons}). Neither enters
$T_{\mu\nu}$, as shown below, so the four configurations share the same geometry. The gluon field lies along a single generator, but this is inessential (Appendix~\ref{app:nonabelian}).

The field equations are checked in turn. In the tetrad $h^I{}_\mu=\delta^I_\mu+\frac{1}{2}Wk^Ik_\mu$ the spinor $\chi$ is
constant, and both the spin connection and $\gamma^\mu A_\mu$ are proportional to $\gamma^\mu k_\mu$, which annihilates $\chi$. Hence
$\mathring{\nabla}_\mu\psi_f=k_\mu\lp n'/2n-i\Phi'+ig_s\lambda_fx^i\varepsilon_i\rp\psi_f$ and the massless Dirac equations hold for any
profiles. The colour current vanishes, $\sum_fc_f^\dagger t_Ac_f=\text{tr}\,t_A=0$, $F$ solves the Yang--Mills equations, and the spin
current, proportional to $\epsilon^{\lambda\mu\nu\rho}k_\rho$, is divergence-free. The $\lambda_f$-dependent quark energies cancel in the
flavour sum, so that
\be \label{einstein}
\kappa\mathring{G}_{\mu\nu}=-\frac{\kappa}{2}\lp\partial_x^2+\partial_y^2\rp W\,k_\mu k_\nu=\lp\rho+E_0^2\rp k_\mu k_\nu=T_{\mu\nu}\,.
\ee
For generic profiles, neither $\partial_t$ nor the rotation $\varphi=x\partial_y-y\partial_x$ is an isometry.

Since $\e^a=\diff x^a$, Eq.~(\ref{superpotential}) with $X=-\mathring{\Gamma}$ and $\sqrt{-g}=1$ gives
\be \label{Hpp}
H_a{}^{\lambda\mu}=\frac{\kappa}{2}k_a\lp k^\lambda\partial^\mu W-k^\mu\partial^\lambda W\rp\,, \quad
\partial_\lambda H_a{}^{\lambda\mu}=\kappa\mathring{G}^\mu{}_a
\ee
for any $W(u,x,y)$. By Eq.~(\ref{JG}) the Cartesian reference is exactly canonical.

\emph{The cell.} Since the waves fill the transverse plane, we consider a finite periodic cell rather than an isolated
wave packet. Take $V=D^2_R\times S^1_L$ (Fig.~\ref{fig:cell}a), with $z\sim z+L$, $L$-periodic profiles,
$\Phi(u+L)-\Phi(u)\in2\pi\mathbb{Z}$, $\omega_gL\in2\pi\mathbb{Z}$ and $1+W>0$. The identification removes the longitudinal
endcaps, leaving the toroidal boundary $r=R$. On the boundary worldtube we fix the wave amplitudes,
while allowing normalized null polarizations $\chi$, rotated gluon and gravitational-wave polarizations, and rigid transverse rotations
of $\e$. No symplectic flux crosses $r=R$, since $\chi_1^\dagger\gamma^{\hat{0}}\gamma^{\hat{r}}\chi_2=0$ for any two null
polarizations, $F^{r\nu}\delta A_\nu=0$ for $\delta A\propto k$ with $k^r=0$, the gravitational symplectic current \cite{Koivisto:2026surface} vanishes for
$\delta h^I{}_\mu\propto k^Ik_\mu$, and the potential of rigid frame rotations vanishes. The
rotation $\varphi$ preserves this family, acting on the gravitational wave as $\delta_\varphi b_+=2b_\times$,
$\delta_\varphi b_\times=-2b_+$ at fixed $b_+^2+b_\times^2$.

\emph{Energy and momentum.} The spin term in Eq.~(\ref{surface}) integrates to zero on the circular boundary, and since $k^r=0$ only
$H_a{}^{rt}=\frac{\kappa}{2}k_a\partial_rW$ contributes. The translation charge is a gravitational Gauss flux,
\ba \label{Pa}
P_a & = & \oint_{\partial V}Q_a=\frac{\kappa}{2}k_a\oint_{\partial V}\partial_rW\,\diff S \nn \\
& = & -k_a\int_V\lp\rho+E_0^2\rp\diff^3x\,,
\ea
so that $E=-P_t=\int_V\lp\rho+E_0^2\rp\diff^3x>0$ is the quark and gluon energy in $V$, with $P_z=E$. The harmonic terms $b_+$ and $b_\times$ drop out, so the gravitational wave carries no
energy-momentum in the canonical frame, whereas in the detector frame of Ref.~\cite{Koivisto:2026gw} it carries the Isaacson energy in the
weak-wave limit \cite{Isaacson:1968zza}.

\emph{Rotation.} Since $k_\mu\varphi^\mu=0$, the term $f^aH_a$ in Eq.~(\ref{Q}) vanishes, and likewise $T^t{}_\nu\varphi^\nu=0$.
The spin term survives although $\partial_\lambda\mathcal{S}^{\lambda\mu\nu}=0$, because $\varphi$ grows linearly in the transverse
coordinates. Equation~(\ref{currentT}) gives the density $J^t_\varphi=\mathcal{S}^{txy}=\frac{1}{2}n\,\chi^\dagger\Sigma_3\chi$. The
rotation charge thus comes entirely from the quark spin, and Eq.~(\ref{hamiltonian}) gives
\be \label{Hphi}
H_\varphi=\oint_{\partial V}Q_\varphi=\frac{N_q}{2}\chi^\dagger\Sigma_3\chi\,, \quad N_q=\pi R^2\int_0^Ln\,\diff z\,,
\ee
for either gluon helicity. Opposite quark helicities give $H_\varphi=\pm N_q/2$, while $g_{\mu\nu}$,
$T_{\mu\nu}$ and the canonical frame are identical (Fig.~\ref{fig:cell}b). The splitting is independent of the reference
frame, since the spin term does not involve it and the gravitational term is the same for both quark helicities.

By periodicity $P_a$ and $N_q$ are constant in $t$, so energy and rotation charges are conserved. Since $T^{\mu\nu}\propto k^\mu k^\nu$ with $k$ covariantly constant, the exchange term in
Eq.~(\ref{balance}) reduces to $k^\mu\partial_\mu(k\cdot\xi)$, which vanishes here, so the matter charges are also conserved
separately.

\emph{Helicity.} Neither gluon nor gravitational-wave helicity enters the bulk charge. A canonical spin density
$\vec{E}\times\vec{A}$ in the transverse gauge would assign the gluons a helicity proportional to $\pm E_0^2/\omega_g$ per unit
volume \cite{Jaffe:1989jz,Leader:2013jra}. Here, as for the Belinfante angular momentum of a plane wave \cite{Ohanian:1986}, gluon spin
can enter only where the field is non-uniform.
The boundary family considered here is narrow, and extending the analysis beyond it is left open.

\section{Conclusion}

Improvement identities relate tensors; they do not by themselves select a physical current. In G$_\parallel$R the variational principle
supplies the currents, and the canonical frame selects which of their surface charges are physical. For QCD this gives the
gauge-invariant tensor and the Belinfante--Rosenfeld relation from the action, energy and rotation charges conserved without the
corresponding isometries, and rotation charges that distinguish quark helicities invisible to the metric and the symmetric tensor.

Two lessons follow. First, a spin current is the response of matter to the Lorentz connection. A gauge field has none, its curvature
(\ref{F}) involving no spacetime connection, so gluon angular momentum is carried by field momentum and localizes where the field
varies \cite{Ohanian:1986}. Accordingly, neither gluon nor gravitational-wave helicity enters the bulk rotation charge. Second,
finite-region charges are physical. Whilst the totals for localized states agree with the standard decompositions, G$_\parallel$R
assigns quark and gluon spin to definite regions, also in the presence of gravity.

\begin{acknowledgments}
This work was supported by the Estonian Research Council grants PRG2608 ``Space-Time-Matter'', CoE TK202 ``Foundations of the
Universe'', TARISTU24-TK3 and TARISTU24-TK10.
\end{acknowledgments}

\appendix

\section{Non-Abelian gluon waves}
\label{app:nonabelian}

The exact solution admits a continuous extension to gauge-inequivalent, noncommuting SU(3) field configurations, without changing
the spacetime geometry or the canonical charges. The results of Sec.~\ref{sec:example} therefore do not rely on the gluon wave
(\ref{sol-gluons}) being effectively Abelian.

Replace (\ref{sol-gluons}) by $A=-x^i\mathcal{E}_i(u)\,\diff u$ with
\ba \label{nonabelian}
\mathcal{E}_x & = & E_0\cos(\omega_gu)\,t_3\,, \nn \\
\mathcal{E}_y & = & \sigma E_0\sin(\omega_gu)\lp\cos\alpha\,t_3+\sin\alpha\,t_1\rp\,,
\ea
where $\sigma=\pm1$ and $\alpha$ is constant. The values $\alpha=0$ and $\alpha=\pi$ give (\ref{sol-gluons}) with helicity $\sigma$ and $-\sigma$.
Since $A\propto k$, $F=k\wedge\mathcal{E}$ with $\mathcal{E}=\mathcal{E}_i\diff x^i$, and $k_\mu F^{\mu\nu}=0$ removes the nonlinear
terms, so the Yang--Mills equations hold as before. The colour matrix $x^i\mathcal{E}_i$ enters $\mathring{\nabla}_\mu\psi_f$ only along
$k_\mu$, and $T_{\mu\nu}$ only through $\text{tr}\,\mathcal{E}_i=0$ and $2\,\text{tr}(\mathcal{E}_i\mathcal{E}_i)=E_0^2$. The metric,
the canonical frame and, since still $\delta A\propto k$, the charges $P_a$ and $H_\varphi$ are unchanged. However,
\be
\lb F_{ux},F_{uy}\rb=\tfrac{i}{2}\,\sigma E_0^2\sin\alpha\,\sin(2\omega_gu)\,t_2
\ee
is gauge covariant, so for $\sin\alpha\neq0$ the configuration is not gauge equivalent to an Abelian one, and the invariant
$2\,\text{tr}(F_{ux}F_{uy})=\tfrac{1}{2}\sigma E_0^2\cos\alpha\,\sin(2\omega_gu)$ distinguishes the values $\alpha\in[0,\pi]$.

\end{document}